# Interior X-ray diffraction tomography with low-resolution exterior information

Zheyuan Zhu[1*], Alexander Katsevich[2], and Shuo Pang[1]

[1]CREOL, The College of Optics & Photonics, University of Central Florida, Orlando, FL, 32816, USA

[2]Department of Mathematics, University of Central Florida, Orlando, FL, 32816, USA

*zyzhu@knights.ucf.edu

**ABSTRACT**

X-ray diffraction tomography (XDT) resolves spatially-variant XRD profiles within macroscopic objects, and provides improved material contrast compared to the conventional transmission-based computed tomography (CT). However, due to the small diffraction cross-section, XDT suffers from long imaging acquisition time, which could take tens of hours for a full scan using a table-top X-ray tube. In medical and industrial imaging applications, oftentimes only the XRD measurement within a region-of-interest (ROI) is required, which, together with the demand to reduce imaging time and radiation dose to the sample, motivates the development of interior XDT systems that scan and reconstruct only an internal region within the sample. The interior problem does not have a unique solution, and a direct inversion on the truncated projection data often leads to large reconstruction errors in ROI. To reduce the truncation artifacts, conventional attenuation-based interior reconstruction problems rely on a known region or piecewise constant constraint within the ROI, which do not apply to all the samples in XDT. Here we propose a quasi-interior XDT scheme that incorporates a small fraction of projection information from the exterior region to assist interior reconstruction. The low-resolution exterior projection data obviates the requirement for prior knowledge on the object, and allows the ROI reconstruction to be performed with the fast, widely-used filtered back-projection algorithm for easy integration into real-time XDT imaging modules.

## 1. Introduction

Due to the high penetration depth and material-specific signatures, X-ray diffraction (XRD) is a powerful tool to probe the material properties at molecular level, such as solving the crystalline structure [1,2], identifying the chemical composition in compounds with high sensitivity [3–5], and more recently, diagnose structural defects in bones, prosthetic implants [6,7], or disease-related anomaly in soft tissues [8,9]. For heterogeneous object on the centimeter scale or larger, resolving the XRD profile of individual sample voxels requires a combination of XRD measurement and computed tomography (CT) [10], which captures the XRD signals under different projections and is termed X-ray diffraction tomography (XDT). However, because of the weak XRD signal, XDT scan is typically conducted with a high-brilliance synchrotron source [2,3,6,11,12], which hinders its adoption in industrial and medical applications. A number of XDT systems based on X-ray tubes have been demonstrated using collimators on either the source side or detector side [13–15], but a full scan on an object of ~20mm in diameter involves tens of hours of imaging time, and delivers a high radiation dose to the sample.

In medical diagnosis or industrial inspection, XDT is ideal for a secondary scan, in which only the material composition within an ROI is desired [16,17]. For these applications, interior XDT limits the projection measurement to the ROI region to reduce the imaging time and excessive radiation dose outside ROI, and provides comparable material specificity as a full XDT scan [18]. Similar to the interior problem in CT, interior XDT also requires prior knowledge about the object, either in the form of a known sub-region [19,20] or piecewise constant constraint [21,22], to stabilize the ROI reconstruction. However, knowing the *in-situ* XRD profile of a small region within the whole sample is not possible in XDT. Piecewise constant constraint could be a feasible prior for interior XDT, yet it is not generally applicable to samples with fine features or with a high scattering density gradient. Based on the non-localized filtered back-projection used in XDT reconstruction, here we introduce a small fraction of projection measurement in the exterior region to stabilize the ROI reconstruction. The combined interior and exterior information provides an alternative approach to the existing interior reconstruction techniques that require prior knowledge of the sample.

## 2. Theory

*2.1 X-ray diffraction tomography*

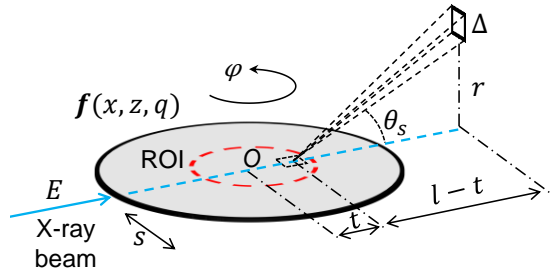

Figure 1: Geometry of pencil-beam X-ray diffraction tomography.

The pencil-beam XDT system (Figure 1) illuminates the sample with a collimated, narrow-band X-ray beam. Upon interacting with the sample, the scattering mechanism of X-ray photons falls into two major categories, Compton incoherent scattering and Rayleigh coherent scattering. In coherent scattering, scattered photons maintain the same energy but are deflected towards a different direction away from the incident direction. Under small scattering angles, coherent scattering is the dominant scattering process [23] that gives rise to the XRD signature of the material. The number of diffracted photons $dI$ in the direction $\theta_s$ from a volume $dV$ is determined by the scattering geometry and molecular form factor

$$dI = I_0 \frac{r_e^2}{2}(1+\cos^2\theta_s)f(\mathbf{r},q)\,d\Omega\,dV\,dq, \tag{1}$$

where $I_0$ is the irradiance of the incident beam; $\theta_s$ is the angle between incident and scattered radiation, and $\cos\theta_s \approx 1$ for small angles; $d\Omega = \Delta^2/(r^2 + (l-t)^2)$ is the solid angle covered by the detector pixel with size $\Delta$ located at distance $r$ away from the pencil beam, and $l$ is the distance between the rotation center and detector plane. The scattering volume $dV = A\,dt$, where $A$ is the cross-section area of the pencil beam; $t$ is the depth of scatter along the pencil beam. $f(x,z,q) = n(x,z)F^2(x,z,q)$ is the product between the density of the scatter $n(x,z)$ and the molecular form factor $F(x,z,q)$. The momentum transfer $q$ that causes the deflection of photons with energy $E$ and scattering angle $\theta_s$ is defined by the Bragg's law

$$q = E/hc \sin(\theta_s/2), \qquad (2)$$

where $h$ and $c$ represent Plank's constant and speed of light, respectively; $E$ is the energy of the incident X-ray.

We can introduce a series of first-order approximations to Eq. (2) that relates XDT to a volumetric tomography reconstruction problem [18]. At small diffraction angles, $\sin(\theta_s/2) \approx r/(2(l-t))$. Given that the distance from the rotation center to the detector, $l$, is much larger than the size of the sample, the Taylor expansion on the $1/(l-t)$ terms around $t = 0$ simplifies the Bragg's law in Eq. (2) to

$$q = \frac{Er}{2hcl}(1 + \frac{t}{l}), \qquad (3)$$

Eq. (3) describes the momentum transfer probed by the detector pixel located at a distance $r$ away from the pencil beam. The intensity on the detector is a multiplexed measurement of the XRD signals along the pencil beam. To resolve the complete object function, $f(x, z, q)$, the sample needs to be translated across the beam by the distance $s$ in the $(x, z)$ plane, and rotated around the vertical axis by the angle $\phi$ similarly to that in a pencil-beam CT. The total number of photons collected by a detector pixel under sample offset $s$ and projection angle $\phi$ is

$$I(r,s)_\phi = Cw(r) \int_\Gamma f(s\hat{\Theta} + t\hat{\Theta}^\perp + \tan\alpha_r \left(1 + \frac{t}{l}\right)\hat{q}) \, dt, \qquad (4)$$

where $C = I_0 A r_e^2$ is a constant proportional to the source intensity; $\hat{\Theta} = (\cos\phi, \sin\phi, 0)$ and $\hat{\Theta}^\perp = (-\sin\phi, \cos\phi, 0)$ are the directional vectors parallel and perpendicular to the incident X-ray beam in the x-z plane, respectively; $\hat{q} = (0,0,1)$ is the unit vector along the momentum transfer dimension; $w(r) = \Delta^2/(r^2 + l^2)$ is a weight to account for the decay of the scattering signal.

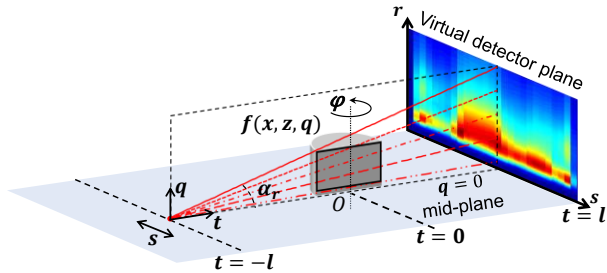

Figure 2: XDT projection as a series of parallel, vertical fan beams in the $(x, z, q)$ domain.

## 2.2 Interior XDT reconstructions

Eq. (4) implies that the integral is performed along the line $\Gamma$ in the $(x, z, q)$ domain with slope $\tan\alpha_r = Er/(2hcl)$, where $\alpha_r$ is the angle with respect to the mid-plane ($q$=0). Figure 2 illustrates a family of XDT projection lines with $r$ as a parameter, which, combined with sample translation, form a series of parallel, vertical fan-beam projections of the object $f(x, z, q)$ onto the virtual detector plane $(s, r)_\phi$ at rotation angle $\phi$. The XDT reconstruction follows a filtered back-projection modified from the FDK-algorithm [24] (modified-FDK, for short), in which the back-projection is performed along diverging lines in the $z$-$q$ plane, and parallel lines in the $x$-$z$ plane.

2.2.1: *Addition of low-resolution exterior projections*: Let $a_1(\phi)$, $a_2(\phi)$ denote the two ends of the object support at each projection angle, such that $I(r,s)_\phi = 0$ if $s < a_1(\phi)$ or $s > a_2(\phi)$. Let $b_1(\phi)$, $b_2(\phi)$ denote the boundary of ROI region, within which the pencil-beam scan is performed at each projection angle. For interior XDT scan, Eq. (4) becomes

$$I(r,s)_\phi = Cw(r)\int_\Gamma f(s\hat{\Theta}+t\hat{\Theta}^\perp + \tan\alpha_r\left(1+\frac{t}{l}\right)\hat{q})\,\mathrm{d}t,\ b_1(\phi) < s < b_2(\phi), \tag{5}$$

which indicates that interior XDT detects the same range of scattering angles and the number of projections as a conventional pencil-beam XDT, except that only the ROI region of the sample is scanned. Analogous to the conversion between fan-beam CT and parallel-beam CT [25], the projection lines in the parallel fan-beam geometry can be rearranged to a half-cone-beam projection of $(x, z, q)$ onto the virtual detector plane $(s, r)_\phi$ at each projection angle. Therefore interior XDT is equivalent to a transversely-truncated cone-beam CT. Inspired by the truncated cone-beam CT reconstruction frameworks [26,27], which combine low-resolution (LR) full field-of-view (FOV) scans with high magnification, ROI-only projections, our quasi-interior XDT introduces additional LR measurements of the exterior region, and linearly interpolate the missing pixels between two adjacent LR samplings to form a complete sinogram with homogenous resolution in both ROI and exterior region. The modified-FDK then back projects the complete sinogram $I(r,s)_\phi$ to the object domain $(x, z, q)$ to reconstruct the full FOV, from which only the ROI region is extracted.

*2.2.2 Extrapolation on truncated ROI projection*: In the filtered back-projection framework, the ROI reconstruction from truncated projection data does not have a unique solution, since any exterior data, combined with the interior projection, can be inversed by modified-FDK. Based on the continuity of the projection data associated with a realistic sample, it is possible to supplement the missing exterior projection via a continuous extrapolation on the truncated projection from ROI boundary to the object support [28]. Specifically, we extrapolate the measured interior sinogram into the exterior region via

$$I(r,s)_\phi = \begin{cases} c_1\sqrt{s-a_1(\phi)},\ s<b_1(\phi) \\ c_2\sqrt{a_2(\phi)-s},\ s>b_2(\phi) \end{cases}, \tag{6}$$

where $c_1, c_2$ are constants that ensures continuity across the ROI boundary, and are calculated according to

$$\begin{cases} c_1 = \dfrac{I(r,b_1(\phi))_\phi}{\sqrt{b_1(\phi)-a_1(\phi)}} \\ c_2 = \dfrac{I(r,b_2(\phi))_\phi}{\sqrt{a_2(\phi)-b_2(\phi)}} \end{cases}. \tag{7}$$

*2.2.3 Total-variance (TV) regularization*: In addition to the filtered back-projection–based frameworks, iterative reconstruction that incorporates total variation (TV) regularization (IR-TV for short) to enforce the piecewise constant constraint is also widely used in interior problem. In the iterative reconstruction algorithm, both object **f** and measurement **I** are discretized. The system matrix **H** relating **f** and **I** is constructed from the pixels hit by each beam and the weight $w(r)$ in Eq. (4). Each pencil-beam measurement is an independent entry in the system matrix **H** in IR-TV. For interior measurement, we limit the size of the matrix **H** to cover only the pencil beams passing through ROI region. The system matrix **H** can be extended to include additional exterior scans without interpolating the LR exterior sampling onto the same, uniform grid as in the interior scan.

IR-TV is an optimization problem that minimizes the objective function in Eq.(8), which consists of a Poisson likelihood $P(\mathbf{I}|\mathbf{f})$ of observing the measurement $\mathbf{I}$ given the object $\mathbf{f}$, and a TV regularizer

$$\hat{\mathbf{f}} = \arg\min_{\hat{\mathbf{f}}'}\{-\log P(\mathbf{I}|\mathbf{f}') + \tau TV(\mathbf{f}')\}. \qquad (8)$$

The TV operator is defined as

$$TV(\mathbf{f}) = \sum_{i,j,k} \sqrt{(f_{i+1,j,k} - f_{i,j,k})^2 + (f_{i,j+1,k} - f_{i,j,k})^2} \qquad (9)$$

in which the index $i$ and $j$ represents the spatial dimension $x$ and $z$, respectively. $k$ represents the index in the momentum transfer domain. The optimization was implemented by an EM algorithm [29] embedded with TV regularization [30] in each iteration. The iteration terminates when the relative change in the objective function (Eq. (8)) is smaller than $10^{-4}$. The parameter $\tau$ that yields the minimal objective function at the end of the iterations is selected.

## 3. Simulation and experiment setup

A numerical phantom was designed to evaluate the performance of our quasi-interior XDT reconstruction based on the modified-FDK, and compare it with existing interior reconstruction methods. Figure 3(a) shows the geometric configuration of the phantom, which consists of circular regions of five different diameters, marked by the digit next to each circle. The smallest to the largest diameters are 0.8, 1.4, 1.8, 2.6 and 3.2mm, respectively. Each region is filled with either water or Nylon, and wrapped in a 26mm-diameter Lucite tube. The tube is filled with saturated fat whose density, $n(x,z)$, follows a Gaussian function along the radial direction. XRD profiles (Figure 3(b)) reported in the previous work [10,31] were assigned to each material region in the phantom. The 10mm-diameter dashed circle in (a) marks the ROI, within which the sampling interval is 0.25mm. Without losing generality, we coincide the center of both the Lucite tube and ROI with the rotation center, so that $a_1$, $a_2$, $b_1$, and $b_2$ are all independent on $\phi$. The LR exterior scan additionally samples 6 pixels separated 2.5mm apart in the exterior region per projection angle. The rotation step size was 1 degree for all the measurements. The XDT measurements were generated by the system matrix $\mathbf{H}$ constructed from the distance-driven projection geometry [32]. For the iterative reconstruction, we constructed three system matrices. $\mathbf{H}_1$ is the full XDT scan with 0.25mm sampling across the whole FOV. $\mathbf{H}_2$ covers only the interior region sampled at 0.25mm, and $\mathbf{H}_3$ adds the 6 exterior pixels per projection angle to $\mathbf{H}_2$. Figure 3(c) shows the sinogram $I(s,\phi)_r$ at $r$=10, 20 and 30mm of the full XDT scan generated by $\mathbf{H}_1$. The interior scan (d) was simulated by extracting the ROI projection from the full scan in (c). Figure 3(e) shows the modified-FDK reconstruction from truncated sinograms, in which the artifacts on the ROI boundary are clearly visible.

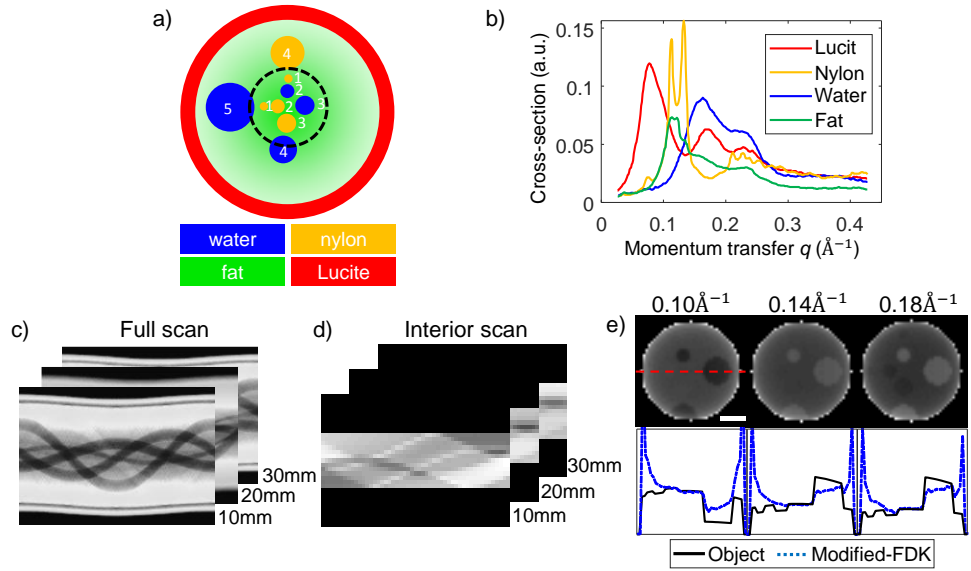

Figure 3: Simulation setup of the interior XDT system. (a) The geometric configuration of the phantom. (b) XRD profile of each material in (a). (c) Full scan XDT sinograms at $r$=10, 20 and 30mm. (d) Interior scan sinogram from truncating the measurement in (c). (e) Modified-FDK reconstruction on the truncated sinogram. The spatial profiles are plotted along the red, dashed line. The scale bar represents 5mm.

The experiment employed a copper-anode X-ray tube (XRT60, Proto Manufacturing) operating at 45kV and 40mA, which emits a characteristic spectrum peak at 8keV. A pair of lead pinholes collimated the beam to a diameter of 2mm. The sample was mounted on a rotational stage (RV1200P, Newport) to cover 180° projection angles with 3.6° step size. The sample consisted of three cylindrical vials filled with water, soybean oil and ethanol, and was wrapped inside a 27mm-diamter, 1mm-thick Teflon tube. A circular region (11mm in diameter) at the center of the sample was marked as the ROI. The sample was translated by a linear stage (UTM150CC, Newport) across the pencil beam with step size of 1mm inside the ROI, and 6mm outside the ROI. The diffracted X-ray was captured by a flat panel detector (1215CF-MP, Rayence) located at 120mm away from the sample. The detector pixel size is 0.2mm under 4X binning model. We performed an azimuthal binning within a series of concentric, 2-pixel-wide rings to reduce each 2D image to 1D intensity profile along the radial direction. The central 10mmX10mm region on the detector was covered by a lead beam stop to block the transmitted beam. The acquisition time for each diffraction pattern was 30 seconds. For the 32mm-diameter phantom, the total imaging time of 15 XRD patterns, 11 in the ROI and 4 in the exterior region, under 50 projections was 6.5 hours. To measure the single-point XRD reference, the 2mm-wide tip of each vial was scanned by the pencil beam to collect 6 XRD profiles for water, oil and ethanol. A CT scan of the whole sample with 0.1mm spatial resolution under 180° projections at 1° step size was also performed.

## 4. Results and discussion

*4.1 Simulation results*

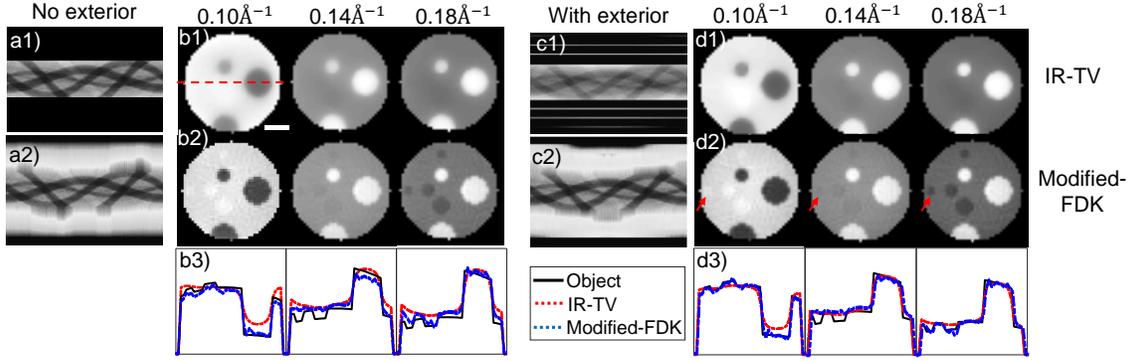

Figure 4: Comparison between interior XDT reconstructions without and with additional low-resolution exterior information. (a) Interior scan sinogram (a1) and extrapolation to the object support (a2), both shown for the sinogram at $r$=10mm. (b) IR-TV reconstruction from interior measurement using $\mathbf{H}_2$. (b2) Modified-FDK reconstruction from extrapolated sinogram. (b3) intensity profile of each image along the red, dash line on (b1). (c) Interior XDT sinogram combined with low-resolution exterior information: (c1) a direct combination between interior and exterior projection for use in IR-TV based on $\mathbf{H}_3$; (c2) interpolated sinogram with homogenous sampling in both interior and exterior region. (d1-d3) ROI reconstruction and spatial profile from (c) using IR-TV and modified-FDK, respectively. The red arrows in (d2) mark the 0.4mm feature. The scale bar in (b1) represents 5mm.

We first reconstructed the ROI region without additional exterior information by enforcing piecewise constant constraint or continuously extrapolating the truncated sinogram. Figure 4(b) compares the IR-TV reconstruction from truncated measurement (a1) using system matrix $\mathbf{H}_2$, and modified-FDK reconstruction from extrapolated sinogram (a2). The normalized mean square error (NMSE) between the reconstructed ROI and ground truth is 4.8% for IR-TV, and 3.5% for modified-FDK with extrapolated sinogram. The reconstructed spatial profiles in Figure 4(b3) indicate that IR-TV and sinogram extrapolation can both suppress the truncation artifacts in interior XDT with comparable reconstruction performance, yet they require prior knowledge of the object, either in the form of piecewise constant for IR-TV, or the precise size of object support for sinogram extrapolation. It is worth noting that for the sinogram extrapolation in (a2), our simulation uses the true support size $a_1(\phi)=-a_2(\phi)$=13mm, which corresponds to the size of the Lucite tube. A wrong value for $a_1$ and $a_2$ will yield less accurate reconstructions, which is demonstrated by the experiments.

Next, we combined the truncated projection with a simulated LR exterior scan in ROI reconstruction. For modified-FDK, the exterior projection was linearly interpolated to match the sampling interval of the ROI (c2). For IR-TV, no interpolation was performed on the combined interior/exterior measurement (c1). Figure 4(d) compares the reconstruction from modified-FDK on the interpolated sinograms and IR-TV using $\mathbf{H}_3$. The NMSE in the ROI is 1.47% for IR-TV, and 1.35% for modified-FDK. Compared to interior reconstruction, both of these two reconstruction methods exhibit reduced reconstruction error when a small fraction (~10%) of exterior information is present. The lower reconstruction error of modified-FDK is attributed to the fine feature (0.4mm region, marked by the red arrows) and the intensity gradient preserved in the reconstructed ROI, both of which are compromised on IR-TV reconstruction.

*4.2 Experiment results*

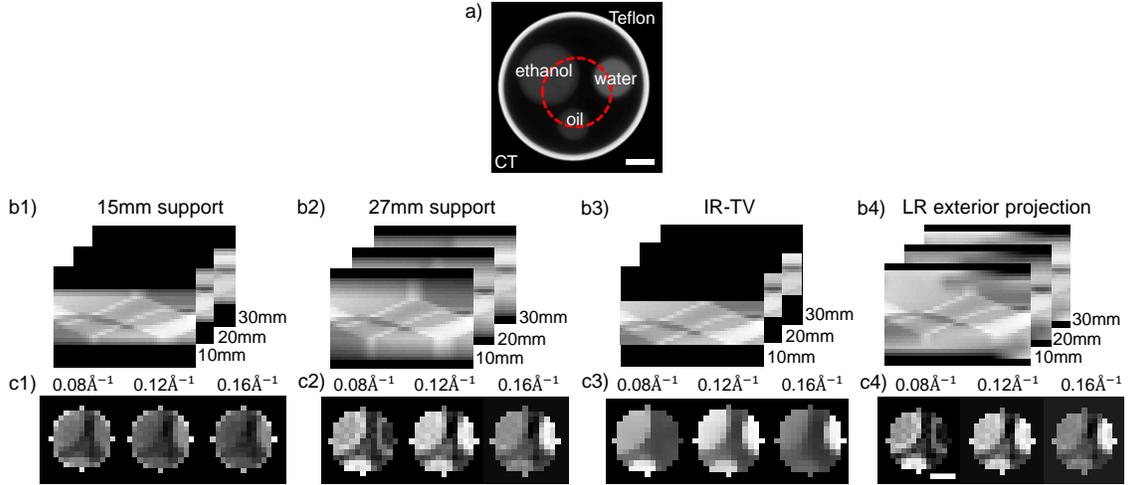

Figure 5: Comparison between the XDT reconstruction with and without exterior information. (a) A CT image of the phantom. The red, dashed line marks the ROI region. (b) XDT sinograms at $r$=10, 20, and 30mm corresponding to: (b1-b2) interior projection extrapolated to a support of 15mm (b1) and 27mm (b2) in diameter, (b3) pure interior projection for IR-TV reconstruction, (b4) interior projection combined with interpolated low-resolution exterior scan. (c) Reconstructions at $q$=0.08, 0.12 and 0.16Å$^{-1}$ from (b1-b4). All reconstructions are performed with modified-FDK except (c3), which is reconstructed with IR-TV. The scale bars in (a) and (c4) both represent 5mm.

A CT image of the phantom analyzed by our system is shown in Figure 5(a). The red, dashed circle on the CT image marks the ROI region for interior XDT scan. The average grayscale value (normalized to maximum) within oil and ethanol regions are 0.17±0.03 and 0.19±0.03, respectively. The contrast between these two materials is comparable with the fluctuation in the reconstructed image, which renders them indistinguishable on CT. XDT reconstructions were carried out using the modified-FDK (b1, b2 and b4) and IR-TV (b3) algorithms. The sampling grid for both two algorithms is 1mm in spatial domain, $(x, z)$, and 0.005Å$^{-1}$ in momentum transfer domain, $q$. Figure 5(b) compares the XDT reconstructions in the ROI region from extrapolation (b1-b2), IR-TV (b3) and LR exterior scan (b4). The sinograms used in the reconstructions are shown for 3 representative $r$ at 10, 20 and 30mm (b1-b4). The reconstructed ROI regions (Figure 5(c)) are displayed under 3 momentum transfer values $q$=0.8, 0.12, and 0.16Å$^{-1}$. Figure 5(b1) extrapolates the truncated sinogram to 15mm-diameter support, which is smaller than the actual size of the object, and its reconstruction (c1) exhibits artifact at ROI boundary, a phenomenon consistent with previous analytical results [33,34]. Figure 5(b2-b3) shows that both IR-TV and a correct extrapolation to the real support size alleviate the truncation artifact without exterior information. Due to the use of TV regularization in (c3), ethanol and water contrast less with the surrounding empty region than those in (c2). Figure 5(c4) displays the modified-FDK reconstruction with interpolated LR exterior data. The average grayscale at 0.08Å$^{-1}$ (normalized to maximum) in Figure 5(c4) is 0.91±0.07 within the oil region, and 0.66±0.08 within the ethanol region, which clearly distinguishes oil and ethanol. The higher intensity of oil for $q$ ranging from 0.06 to 0.12Å$^{-1}$ agrees with its larger molecular form factor [35].

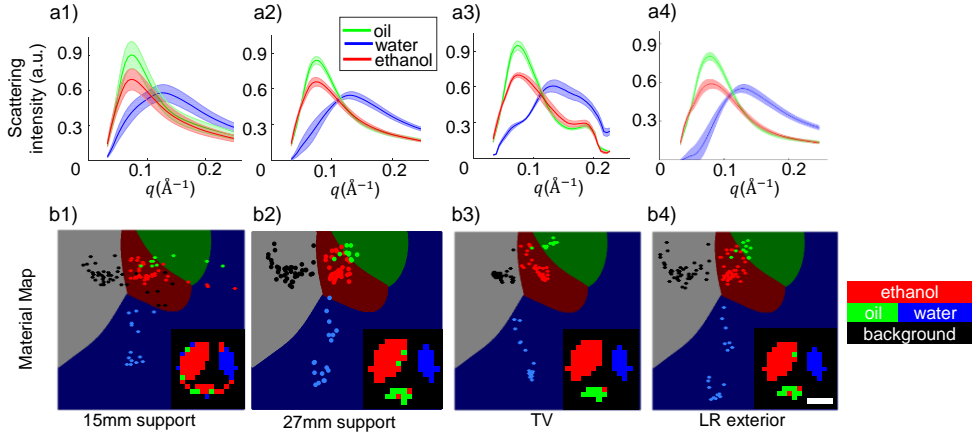

Figure 6: Material classification based on the reconstructed XRD profile with and without exterior information. (a1, a2) Mean and standard deviation of the reconstructed XRD profiles for each material within ROI. (b1, b2) SVM classifier on the principal coordinates for oil, water, ethanol and background. The inset shows the material map of the ROI. Scale bar represents 5mm.

Figure 6(a) plots the mean and standard deviation of the reconstructed XRD profiles within water, ethanol and oil regions. The primary scattering peaks measured by our system are located at $0.12\text{Å}^{-1}$ for water, and $0.077\text{Å}^{-1}$ for oil, which are both lower than previous XRD measurements [36] by a factor of 0.7 due to the beam-hardening effect on the low-energy spectrum, a phenomenon that is also observed in [31,37]. The XRD profiles reconstructed from extrapolation, IR-TV and LR exterior scan are compared in (a1-a4). The reconstructed profiles from wrong sinogram extrapolation (Figure 6(a1)) display high uncertainty, and the water profile exhibits an abnormal shoulder at $0.08\text{Å}^{-1}$. The proximity of this $0.08\text{Å}^{-1}$ shoulder to the peak location of oil and ethanol signifies the overlapping between XRD profiles reconstructed from wrong support size.

To quantify and compare the material specificity between two interior XDT schemes, material classification has been performed using support vector machines (SVMs) [38] with Gaussian kernel. The reference XRD measurements of each material define the boundary of each material classifier, as shown in Figure 6(b). For illustration purposes, the XRD profile of every pixel in the ROI is represented by a dot in the two-dimensional space spanned by the first two principal components ($\xi_1, \xi_2$) obtained from the principal component analysis. To separate the background from three material classes, 10 randomly-sampled pixels in the gap between two materials were also included in the training set. The insets in Figure 6(b1-b4) show the material map of the ROI in each scenario. Among all the sampling points in ROI region, 16%, 3.5%, 0.8%, and 2.6% are misclassified in (b1-b4), respectively, which clearly indicates that a wrong sinogram extrapolation leads to higher classification error. The material maps of correct sinogram extrapolation, IR-TV and LR exterior scan (b2-b4) have similar performance, which agrees with the previously reported results [18] that IR-TV can achieve comparable material specificity as a full-scan XDT. In Figure 5 and Figure 6, a correctly extrapolated sinogram and our proposed LR exterior scan display similar reconstruction results and material specificity. Despite the small difference in their performances, it is worth noticing that in real XDT applications, the exact support of the object under every projection angle is difficult to obtain without a full-FOV pre-diagnostic CT image, which also subjects the sample to additional scans.

## 5. Conclusion

In summary, we have demonstrated a quasi-interior XDT system that combines a low-resolution exterior scan to reconstruct the ROI region from limited projection information. Compared to the existing interior tomography methods that rely either on the precise size of the object support or piecewise constant constraint, our LR exterior scan obviates the prior in the reconstruction, and thus can be applied to reconstruct a wider variety of samples, especially for samples with fine details or density gradient as demonstrated in the simulation. Although IR-TV, sinogram extrapolation and LR exterior scan all suppress the truncation artifacts and provide good material discrimination based on the reconstructed XRD profile in our experiments, the proposed method does not require knowing the exact size of the object support, or compromise the contrast among different regions in the sample. In addition, the modified-FDK algorithm in our quasi-interior XDT framework is faster and requires less memory than iterative algorithms. The increased reconstruction speed and the reduced demand on computational resources could enable our prototype as an embedded, real-time reconstruction module on XDT scanners.

**Acknowledgement**

The work is supported by National Science Foundation (DMS-1615124).

**Reference**

1. F. Schaff, M. Bech, P. Zaslansky, C. Jud, M. Liebi, M. Guizar-Sicairos, and F. Pfeiffer, "Six-dimensional real and reciprocal space small-angle X-ray scattering tomography," Nature **527**, 353–356 (2015).

2. C. G. Schroer, M. Kuhlmann, S. V. Roth, R. Gehrke, N. Stribeck, A. Almendarez-Camarillo, and B. Lengeler, "Mapping the local nanostructure inside a specimen by tomographic small-angle x-ray scattering," Appl. Phys. Lett. **88**, 1–4 (2006).

3. P. Bleuet, E. Welcomme, E. Dooryhée, J. Susini, J.-L. Hodeau, and P. Walter, "Probing the structure of heterogeneous diluted materials by diffraction tomography.," Nat. Mater. **7**, 468–72 (2008).

4. L. Valentini, M. C. Dalconi, M. Parisatto, G. Cruciani, and G. Artioli, "Towards three-dimensional quantitative reconstruction of cement microstructure by X-ray diffraction microtomography," J. Appl. Crystallogr. **44**, 272–280 (2011).

5. J. Sottmann, M. Di Michiel, H. Fjellvåg, L. Malavasi, S. Margadonna, P. Vajeeston, G. B. M. Vaughan, and D. S. Wragg, "Chemical Structures of Specific Sodium Ion Battery Components Determined by Operando Pair Distribution Function and X-ray Diffraction Computed Tomography," Angew. Chemie - Int. Ed. **56**, 11385–11389 (2017).

6. S. R. Stock, F. De Carlo, and J. D. Almer, "High energy X-ray scattering tomography applied to bone," J. Struct. Biol. **161**, 144–150 (2008).

7. C. Mochales, A. Maerten, A. Rack, P. Cloetens, W. D. Mueller, P. Zaslansky, and C. Fleck, "Monoclinic phase transformations of zirconia-based dental prostheses, induced by clinically practised surface manipulations," Acta Biomater. **7**, 2994–3002 (2011).

8. F. He, A. E. Chiou, H. C. Loh, M. Lynch, B. R. Seo, Y. H. Song, M. J. Lee, R. Hoerth, E. L. Bortel, B. M. Willie, G. N. Duda, L. A. Estroff, A. Masic, W. Wagermaier, P. Fratzl, and C. Fischbach, "Multiscale characterization of the mineral phase at skeletal sites of breast cancer metastasis," Proc. Natl. Acad. Sci. 201708161 (2017).

9. R. M. Moss, A. S. Amin, C. Crews, C. A. Purdie, L. B. Jordan, F. Iacoviello, A. Evans, R.

D. Speller, and S. J. Vinnicombe, "Correlation of X-ray diffraction signatures of breast tissue and their histopathological classification," Sci. Rep. **7**, 12998 (2017).

10. G. Harding, J. Kosanetzky, and U. Neitzel, "X-Ray-Diffraction Computed-Tomography," Med. Phys. **14**, 515–525 (1987).

11. T. H. Jensen, M. Bech, O. Bunk, M. Thomsen, A. Menzel, A. Bouchet, G. Le Duc, R. Feidenhans'l, and F. Pfeiffer, "Brain tumor imaging using small-angle x-ray scattering tomography.," Phys. Med. Biol. **56**, 1717–1726 (2011).

12. U. Kleuker, P. Suortti, W. Weyrich, and P. Spanne, "Feasibility study of x-ray diffraction computed tomography for medical imaging," Phys. Med. Biol. **43**, 2911 (1998).

13. M. S. Westmore, A. Fenster, and I. A. Cunningham, "Tomographic imaging of the angular-dependent coherent-scatter cross section," Med. Phys. **24**, 3–10 (1997).

14. J. Delfs and J. P. Schlomka, "Energy-dispersive coherent scatter computed tomography," Appl. Phys. Lett. **88**, 10–13 (2006).

15. G. Harding, "X-ray diffraction imaging—A multi-generational perspective," Appl. Radiat. Isot. **67**, 287–295 (2009).

16. G. Harding, "X-ray scatter tomography for explosives detection," Radiat. Phys. Chem. **71**, 869–881 (2004).

17. G. L. Harding and B. Schreiber, "Coherent X-ray scatter imaging and its applications in biomedical science and industry," Radiat. Phys. Chem. **56**, 229–245 (1999).

18. Z. Zhu, A. Katsevich, A. J. Kapadia, J. A. Greenberg, and S. Pang, "X-ray diffraction tomography with limited projection information," Sci. Rep. **8**, 522 (2018).

19. H. Kudo, M. Courdurier, F. Noo, and M. Defrise, "Tiny a priori knowledge solves the interior problem," IEEE Nucl. Sci. Symp. Conf. Rec. **6**, 4068–4075 (2007).

20. X. Jin, A. Katsevich, H. Yu, G. Wang, L. Li, and Z. Chen, "Interior Tomography With Continuous Singular Value Decomposition," IEEE Trans. Med. Imaging **31**, 2108–2119 (2012).

21. G. Wang and H. Yu, "Compressed sensing based interior tomography," Phys. Med. Biol. **54**, 2791 (2009).

22. J. Yang, H. Yu, M. Jiang, and G. Wang, "High Order Total Variation Minimization for Interior Tomography.," Inverse Probl. **26**, 350131–3501329 (2010).

23. P. C. Johns and M. J. Yaffe, "Coherent scatter in diagnostic radiology," Med. Phys. **10**, 40–50 (1983).

24. L. A. Feldkamp, L. C. Davis, and J. W. Kress, "Practical Cone-Beam Algorithm," J. Opt. Soc. Am. a-Optics Image Sci. Vis. **1**, 612–619 (1984).

25. S. A. Lavrov and E. N. Simonov, "Effect of Regrouping of Projecting Data from Fan to Parallel Geometry in Reconstruction of Tomographic Images," Biomed. Eng. (NY). **44**, 114–120 (2010).

26. G. Van Gompel, G. Tisson, D. Van Dyck, and J. Sijbers, "A new algorithm for 2D region of interest tomography," in *Proceedings of SPIE*, J. M. Fitzpatrick and M. Sonka, eds. (2004), p. 2105.

27. X. Xiao, F. De Carlo, and S. Stock, "Practical error estimation in zoom-in and truncated tomography reconstructions," Rev. Sci. Instrum. **78**, (2007).

28. G. Van Gompel, M. Defrise, and D. Van Dyck, "Elliptical extrapolation of truncated 2D CT projections using Helgason-Ludwig consistency conditions," in *SPIE Medical Imaging*, M. J. Flynn and J. Hsieh, eds. (2006), p. 61424B.

29. T. K. Moon, "The expectation-maximization algorithm," IEEE Signal Process. Mag. **13**, 47–60 (1996).

30. A. Chambolle, "An Algorithm for Total Variation Minimization and Applications," J. Math. Imaging Vis. **20**, 89–97 (2004).

31. G. Kidane, R. D. Speller, G. J. Royle, and A. M. Hanby, "X-ray scatter signatures for normal and neoplastic breast tissues," Phys. Med. Biol. **44**, 1791 (1999).

32. R. L. Siddon, "Fast calculation of the exact radiological path for a three-dimensional CT array," Med. Phys. **12**, 252–255 (1985).

33. G. Wang and H. Yu, "The meaning of interior tomography," Phys. Med. Biol. **58**, R161 (2013).

34. F. Natterer, *The Mathematics of Computerized Tomography* (Siam, 1986).

35. C. T. Chantler, "Theoretical Form Factor, Attenuation, and Scattering Tabulation for Z=1–92 from E=1–10 eV to E=0.4–1.0 MeV," J. Phys. Chem. Ref. Data **24**, 71–643 (1995).

36. J. Kosanetzky, B. Knoerr, G. Harding, and U. Neitzel, "X-ray diffraction measurements of some plastic materials and body tissues," Med. Phys. **14**, 526–532 (1987).

37. Z. Zhu and S. Pang, "Three-dimensional reciprocal space x-ray coherent scattering tomography of two-dimensional object," Med. Phys. **45**, 1654–1661 (2018).

38. C. Cortes and V. Vapnik, "Support-Vector Networks," Mach. Learn. **20**, 273–297 (1995).